# Web Mining Techniques in E-Commerce Applications

Ahmad Tasnim Siddiqui
College of Computers and Information Technology
Taif University
Taif, Kingdom of Saudi Arabia

Sultan Aljahdali
College of Computers and Information Technology
Taif University
Taif, Kingdom of Saudi Arabia

## ABSTRACT
Today web is the best medium of communication in modern business. Many companies are redefining their business strategies to improve the business output. Business over internet provides the opportunity to customers and partners where their products and specific business can be found. Nowadays online business breaks the barrier of time and space as compared to the physical office. Big companies around the world are realizing that e-commerce is not just buying and selling over Internet, rather it improves the efficiency to compete with other giants in the market. For this purpose data mining sometimes called as knowledge discovery is used. Web mining is data mining technique that is applied to the WWW. There are vast quantities of information available over the Internet.

## 1. INTRODUCTION
The web is becoming much accepted over the last decade, bringing a strong platform for information distribution, retrieval and analysis of information. These days the web is much popular for a large data repository containing a broad variety of data and knowledge base, in which information are hidden [2]?

Users face problems due to the huge volume of information that is consistently growing. In particular, Web users have issues in getting the correct information due to low precision and low recall page. For example, if a user wants to get any information by using Google and other search engines, it will provide not only Web contents dealing with this topic, but a series of irrelevant information, so called noise pages, resulting in difficulties for users in obtaining necessary information [2].

All these pose a challenge to researchers to discover the web management methods and effective extraction of information from the web. To understand the web mining we should know all about the data mining techniques available. Figure 1 provides taxonomy of web mining:

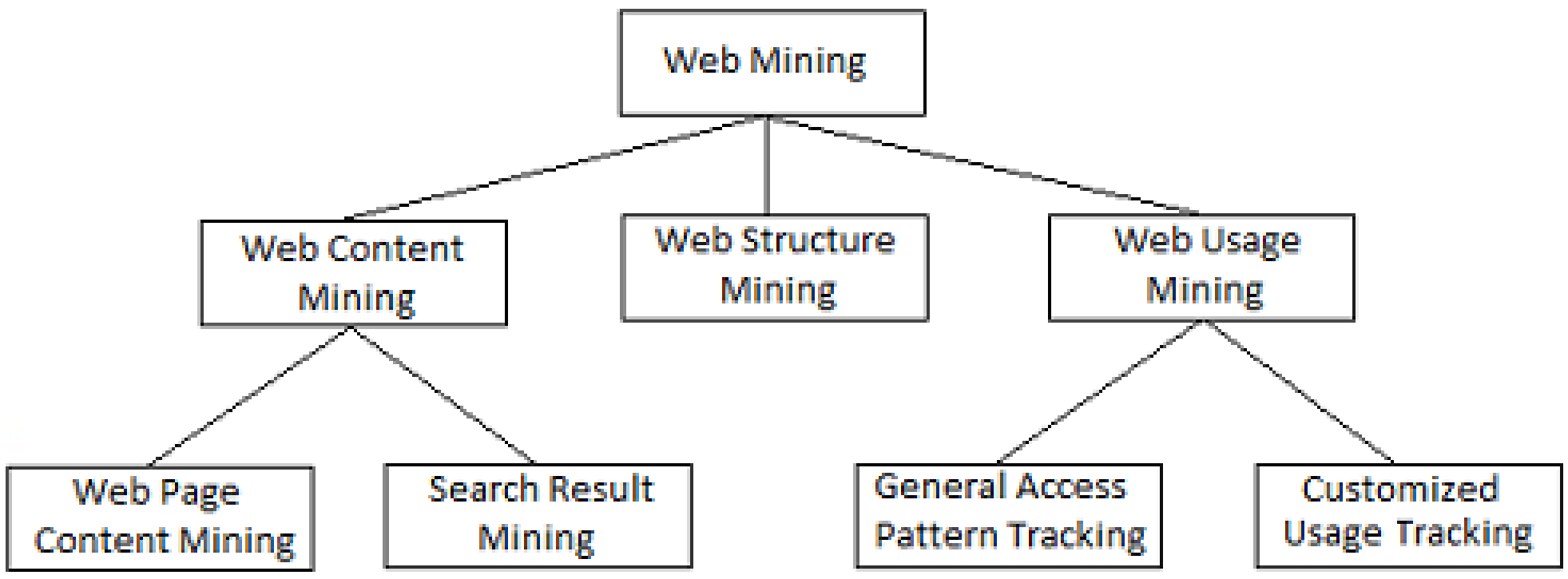

**Fig 1: Taxonomy of Web Mining**

The Web is a critical channel of communication and promoting a company image. The research aims to discover Web information from massive sources of data using data mining techniques. The research on the Web deals with representations of semi-structured and heterogeneous data, such as textual information, hyperlink structure and receives information of use to improve the quality of services provided by different Web applications. These kinds of applications cover a wide category of subjects, such as obtaining the desired Web content, retrieval and analysis of Web communities, user profiles, customized Web presentations in accordance with user preferences [2]. E-commerce sites are important sales channels. E-commerce has changed the face of most business functions in competitive enterprises. It is very important to use data mining methods to analyze data from the activities carried out by visitors to these websites.

In general, e-commerce and e-business have enabled on-line transactions and generating large-scale real-time data has never been easier.

Web content mining is the process of extracting knowledge from documents and content description. Web structure mining is the process of obtaining knowledge from the organization of the Web and the links between Web pages. Figure 2 gives an idea about Web mining research which is divided into three categories: (i) mining the Web content (Web content mining), (ii) mining the web structure (web structure mining) and (iii) mining the Web use (Web usage mining).

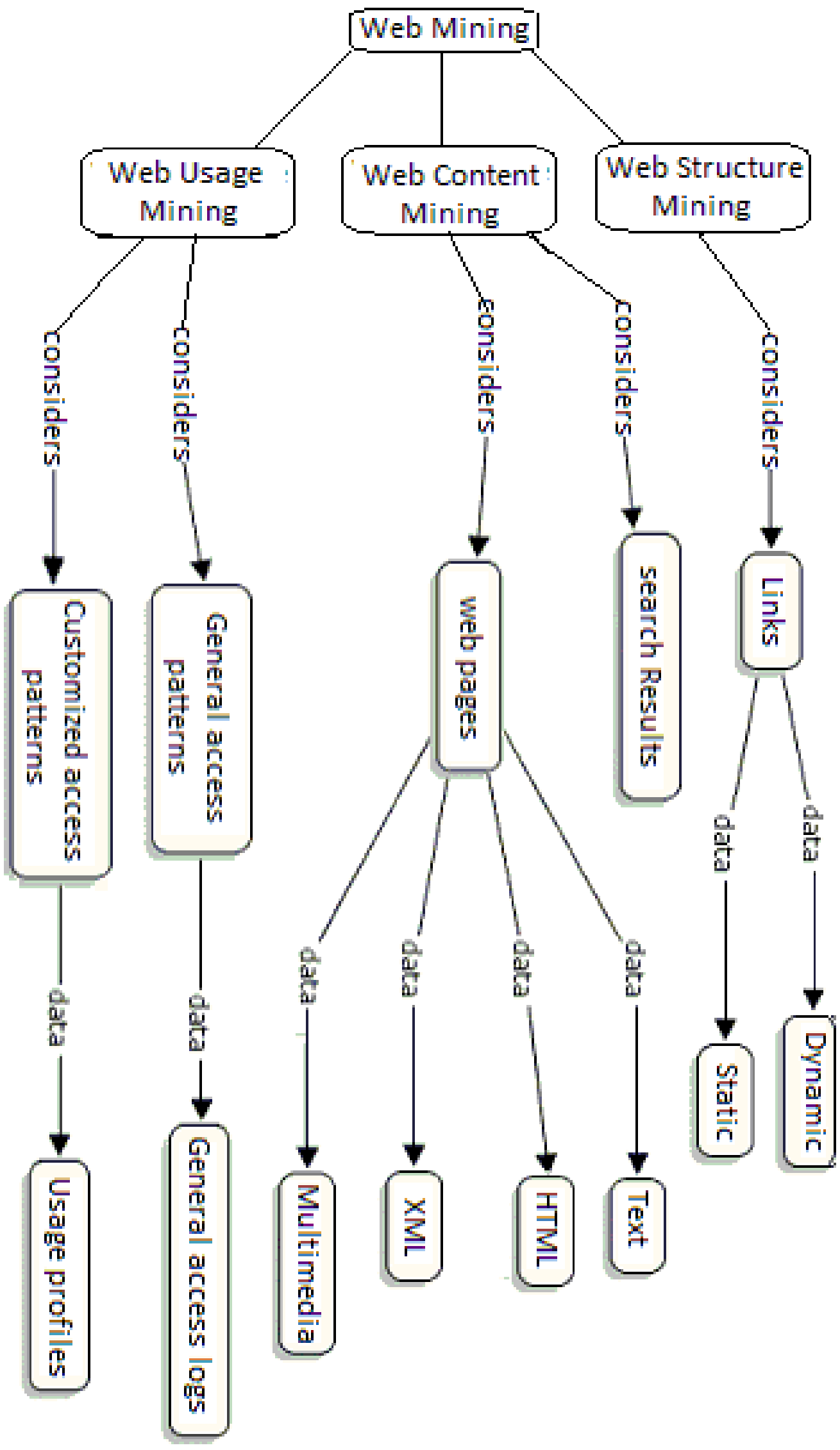

**Fig 2: Web Mining Explained As Tree Structure**

Web usage mining analyzes information about visited Web page that saved in log files of Internet servers in order to discover interesting patterns previously unknown and potentially useful. Web usage is described as mining applying data mining techniques on Web access logs to optimize web site for users.

## 2. OUR CONTRIBUTION

So the final goal of any Data mining implementation in e-commerce is to improve the process work flow that contributes to delivering value to the consumers.

This paper proposes a system model which is very useful in e-commerce applications. This system involves integration of web mining techniques with an e-commerce application. This integration facilitate e-store owner to improve the features and services and also it will help to get information about the customer's or consumer's behavior of visiting web products and services. There are many areas where data mining can be very helpful when integrating with e-commerce. Some of them are: Data mining in customer profiling (Customer profiling means searching for the data which is collected from existing customers of an business organization for patterns that will allow that business organization to predict about who are the potential customers are and how those customers are behaving), Data mining in recommendation systems, Data mining in web personalization, Data mining and multimedia e-commerce, Data mining and behavior of consumer in e-commerce.

## 3. LITERATURE REVIEW

### 3.1 Web Content Mining

In the past few years, the development of the World Wide Web exceeded all expectations. Today, there are several billions of HTML documents, pictures and other multimedia files available via the internet and the number is still rising. Retrieving the interesting contents has become a very difficult task taking into consideration the impressive variety of the Web. Web content consists of several types of data such as text data, images, audio or video, structured records such as lists or tables and hyperlinks. Web content mining can be defined as the scanning and mining of text, graphs and pictures from a Web page to find out the significance of the content to the search query [21].

### 3.2 Web Structure Mining (Web Linkage Mining)

Web structure mining has a lot of challenges to deal with the structure of the hyperlinks within the Web itself. Link analysis is very old area of research. However, with the rising interest in Web mining, the research of structure analysis had improved and these hard works had resulted in a newly emerging research area known as Link Mining [17], which is located at the intersection of the work in link analysis, hypertext and web mining, relational learning and inductive logic programming, and graph mining. There is a potentially wide range of application areas for this new area of research, including Internet.

The Web contains a variety of objects with almost no unifying structure, with differences in the authoring style and content much greater than in traditional collections of text documents. The objects in the WWW are web pages, and links are in-, out- and co-citation (two pages that are both linked to by the same page). Attributes include HTML tags, word appearances and anchor texts [17]. This diversity of objects creates new problems and challenges, since is not possible to directly made use of existing techniques such as from database management or information retrieval. Web Linkage conveys the information how various pages are associated together and formed the huge Web [18].

In the web structure mining we can have the following directions:

- Based on the hyperlinks, categorizing the Web pages and generated the information.
- Discovering the structure of Web document itself.
- Discovering the nature of the hierarchy or network of hyperlinks in the Website of a particular domain.

Web structure mining describes the organization of the content of the web where structure is defined by hyperlinks between pages and HTML formatting commands within a page.

In 1998, two most important hyperlink based algorithms were designed: PageRank and HITS. Both PageRank [2] and HITS [3] were originated from the social network analysis. They exploit the hyperlink structure of the Web to rank pages according to their degree of prestige or authority.

Page Rank algorithm was created in 1997-1998. The most successful Internet search engine, Google works based on this algorithm. Page Rank is rooted in social network analysis and is basically providing a ranking of each web page depending on how many links from other Web pages lead to that page.

Page Rank algorithm is the most commonly used algorithm for ranking the various pages. The Page Rank algorithm is based on the concepts that if a page contains important links towards it then the links of this page towards the other page are also to be considered as important pages. The Page Rank considers the back link in deciding the rank score. If the addition of the all the ranks of the back links is large then the page then it is provided a large rank [1][6][7][8]. If Google were ever going to rank all of the websites in its index in a single list, PageRank would be the sort-by column of that list. It's just a number. A probability calculated from the incoming links to a page. The Page Rank formula is:

$$\text{Page Rank of web site} = \sum \frac{\text{PageRank of inbound link}}{\text{Number of links on that page}}$$

OR

$$PR(u) = (1-d) + d \times \sum \frac{PR(v)}{N(v)}$$

**Fig 3: Page Rank Formulae [22]**

Where, the Page Rank value for a web page u is dependent on the Page Rank values for each web page v, divided by the number N (v) of links from page v.

## 3.3 Web Usage Mining

Web usage mining is the most relevant part in terms of marketing because it explores ways to navigate and conduct during a visit to the website of a company. Methods for extracting association rules are useful for obtaining correlations between the various pages visited during a browsing session. Sequential association rules and time series models can be used to analyze used data from a Web site taking in account a temporal dynamics using the site. Web usage mining is mainly based on sequence analysis of pages visited during a given session, analyzing web clicks. Information about the purchasing behavior of visitors can be taken from the e-commerce sites by analyzing web clicks.

In web usage mining, there is analyzed information on web pages visits that are saved in log files of Internet servers in order to discover the previously unknown and potentially interesting useful patterns. Web usage mining is described as applying data mining techniques on Web access logs to optimize web site for user's interest.

The click-stream means a sequence of Web pages viewed by a user; pages are displayed one by one at a time. When a visitor accesses a website, the server retains all the actions taken by visitors in a log file. A user session describes the sequence of pages viewed by the user during a period of logging on the web pages from several sites.

Each click of the mouse corresponds to a web page request; the sequence of clicks corresponds to such sequence links.

Clickstream or path is click paths mouse clicks that each visitor makes while browsing a site. Initially, sequence analysis of Clickstream determine from where it came the site visitors, the path followed in the web pages for which the analysis is done, time spent on each page, and when and where the visitor travels after leaving the session of the site.

Overall statistics using aggregate information about Clickstream helps us generate analytical reports such as frequency of visits and time that people spend on a particular site and their activity, and frequency of visits and turning those visits into commercial activities.

Analysis of web clicks is the method of gathering and analyzing data about how it crosses the site visitors. Therefore, Clickstream is the order in which the people visit web pages.

Basically, there are two types of analysis of clicks, namely:

- Traffic analysis
- E-commerce analysis

Traffic analysis involves analyzing road traffic that the user follows the site which is based on data collected at the server level analysis using data from clicks. Traffic analysis also records the number of times users loaded pages while browsing.

Analysis of e-commerce use Clickstream data to determine the marketing effectiveness of the site by quantifying user behavior while actually visiting the site visitor browsing the site recording the translation in sales transactions. Analysis of ecommerce is an indicator of the degree of user convenience in using the interface forms, shopping cart, payment etc.

Clickstream analysis has become a very important aspect of online business and advertising. Because they record how users move from one page to another, it can be seen how potential customers have come to these pages and if it stays on the pages of this electronic business or go on. By making the user profile for a specific site, Clickstream analysis has become a type of analysis used for a given Web site which allows another site design to improve customer satisfaction. Analysis of clicks is important for electronic business and to determine the ideal places for advertising. As specified in the article Clickstream Analysis: Both a Business and an Aid in Advertising, Stephanie Mattingly [16] clicks analysis is a significant help in advertising. This analysis can determine the source pages from where it was reached the current web site, and whether the advertising campaigns are running successful or not.

It can be determined if there are other sites with links to current pages, which are the sites and the number of visitors which were directed to the current due to these links. Clickstream analysis can show whether clicks ads on a site run out of this site so instead of leads customers; they direct users to other places to buy. The main purpose in advertising help is to show a connection between the ads that are placed on the Internet and purchases made on website. The analysis also helps to improve click website by customizing it.

# 4. PROPOSED SYSTEM MODEL

The proposed system model to integrate web mining with e-commerce applications shown in the figure 4:

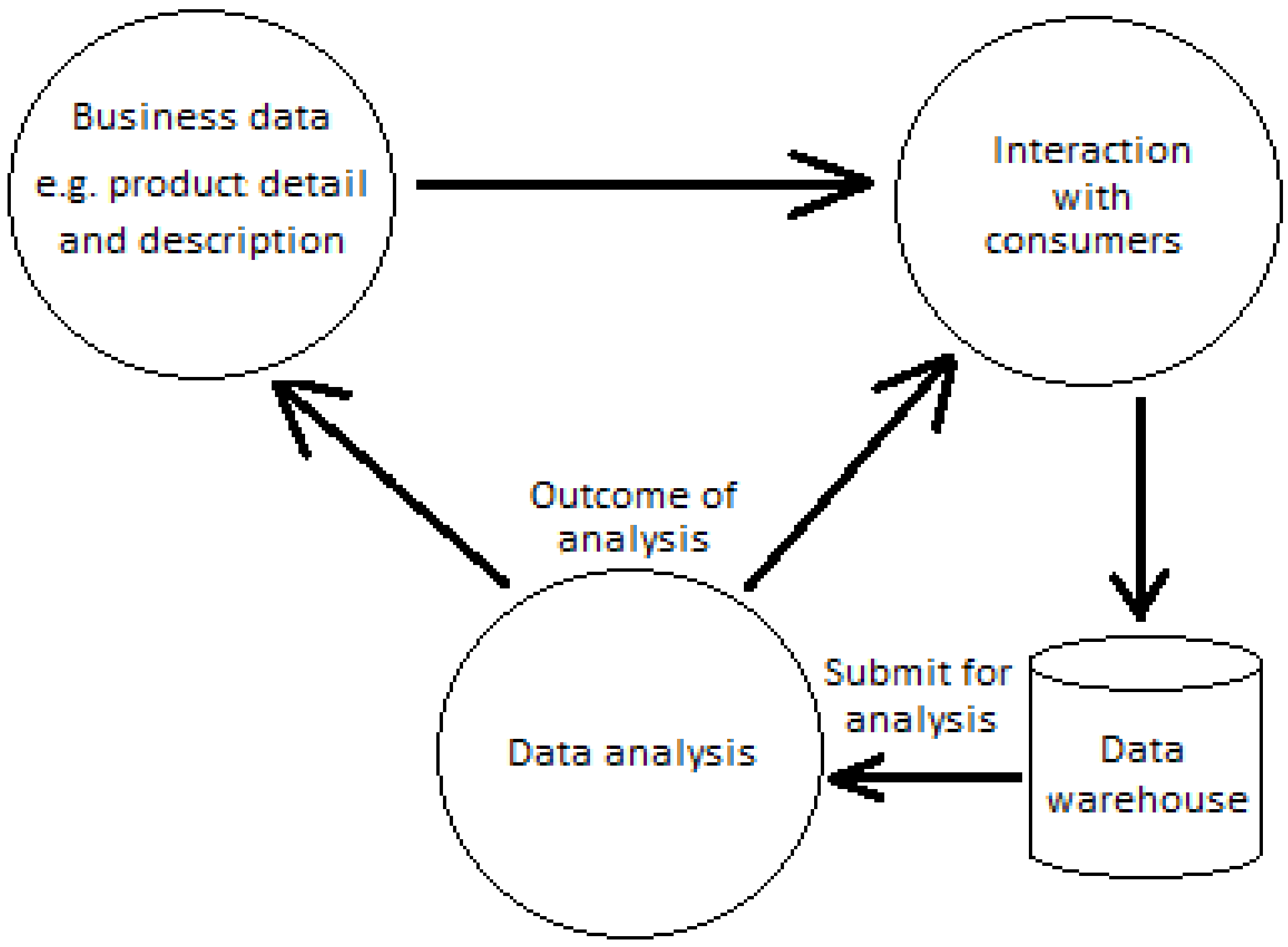

**Fig 4: System model to integrate web mining in E-commerce Applications**

The model shown in figure 4, consists of important components like business data, data obtained from consumer's interaction, data warehouse and data analysis. In case of business data it means the data obtained from the business entities. It may be product & service details and the description. It can provide some important metadata about the product and services. Business data can be collected from RSS feed of particular website or we can read the contents of web page and then data can be extracted from the read contents. In .Net technology there are HttpWebRequest and HttpWebResponse classes in Base Class Library which makes it very easy to request the web pages. Request can be executed like below code:

```
HttpWebRequest request =
(HttpWebRequest)WebRequest.Create("http://www.mypage.com")

.

.

.

HttpWebResponse response =
(HttpWebResponse)request.GetResponse();
```

and after the request data can be read via the response stream code given below:

```
Stream resStream = response.GetResponseStream();
```

From the streaming we can extract the desired data. Second element is interaction with consumer which provides details about the consumer and their choices, there visiting styles, their preferences, forwarding of products and services to friends, number of hits and clicks on particular page and link, nature of consumer and many more. All these information can be collected and inserted into the database or data warehouse. If required, data can also be collected from product ratings page which prompts to the customer to rate the particular items which they recently purchased. These ratings data can be used as input to an analysis engine to help the consumer to find other items that they likely to like. Another element is a data warehouse where we'll keep entire collected data from consumer and business entity for the analysis purpose. Data warehouse is a huge repository for storing the data from business entities and from the consumers for the analysis purpose. After collection of data we'll submit it to analyze the consumer's action. This module is responsible for analysis, decision support, reporting etc. In SQL Server there is a tool Predictive Analytics which provides a highly advanced data mining solution combined with the simplicity and familiarity with MS-Excel [19]. Oracle also provides Data Mining solutions. In Oracle it embeds data mining within the Oracle database [20]. After analysis of entire data the report can be produced to the consumer that the particular how many times product is ordered, popularity of any product and services, best available choice for the consumer etc.

## 5. CONCLUSIONS

In the proposed model web mining integrated with the electronic commerce application to improve the performance of e-commerce applications. First we have discussed some important mining techniques which are used in data mining. After that we explained the proposed architecture which contains mainly four components business data, data obtained from consumer's interaction, data warehouse and data analysis. After finishing the task by data analysis module it'll produce report which can be utilized by the consumers as well as the e-commerce application owners. In future this model can be improved more users interactive and applicable in peer to peer applications [24, 25, 26].

The paper also recommend to work on the semantic web and domain ontology for the purpose of site designing, creation and content delivery semantic web mining was first given by Berendt et al [23]. In future there is requirement to effectively integrate the semantic knowledge from domain ontology which should be able to deal with complex semantic objects.